\documentclass[12pt]{article}
\usepackage{amsmath}%
\usepackage{amsfonts}%
\usepackage{amssymb}%
\usepackage{graphicx}
\begin{document}
\title{Metastable States in Two-Lane Traffic Flow Models With Slow-To-Start Rule}
\author{Najem Moussa $^{1,2}$ \thanks{e-mail: najemmoussa@yahoo.fr}
\\ $^{1}$Facult\'{e} Polydisciplinaire, El Jadida, Morocco
\\ $^{2}$LPMC, Facult\'{e} des Sciences, El Jadida, Morocco}
 \maketitle
\begin{abstract}
Using computer simulations, we show that metastable states still
occur in two-lane traffic models with slow to start rules.
However, these metastable states no longer exist in systems where
aggressive drivers (\textit{which do not look back before changing
lanes}) are present. Indeed, the presence of only one aggressive
driver in the circuit, triggers the breakdown of the high flow
states. In these systems, the steady state is unique and its
relaxation dynamics should depend on the lane changing probability
$p_{ch}$ and the number of aggressive drivers present in the
circuit. It is found also that the relaxation time $\tau$ diverges
as the form of a power-law : $\tau\propto p_{ch}^{-\beta},
\beta=1$.
\newline\
\newline\ PACS. 02.50.-Ey Stochastic processes – 05.45.-a Nonlinear
dynamics and nonlinear dynamic systems – 45.70.Vn Granular models
of complex systems; traffic flow – 89.40.+k Transportation
\end{abstract}
\newpage\
Recently, cellular automata (CA) traffic models are used
enormously in order to understand the complex dynamic behavior of
the traffic in roadways (see the review \cite{chow}). In CA, time
and space are discrete. The space is represented as a uniform
lattice of cells with finite number of states, subject to a
uniform set of rules, which drives the behavior of the system.
These rules compute the state of a particular cell as a function
of its previous state and the state of the neighboring cells. The
most popular CA model for traffic flow on one-lane roadway is the
NaSch model \cite{ns}. Despite its simplicity, the model is
capable of capturing some essential features observed in realistic
traffic like density waves or spontaneous formation of traffic
jams. To describe more complex situations such as multi-lane
traffic, extensions of the NaSch model have been proposed where
additional rules are added for lane changing cars.
\newline\ Barlovic \textit{et al} \cite{Bar} found metastable states in their velocity
dependent randomization (VDR) which is an extension of the NaSch
model. The one-lane VDR model belongs to the class of CA models
with ''slow-to-start'' rules. These models show an hysteresis
effect which is a consequence of the non-unique dependence of the
flow on the density. The above characteristic behavior of traffic
flow is also observed in two-lane traffic models. Indeed, Awazu
\cite{awazu} showed the appearance of several branches and
hysteresis in the relation between traffic flow and car density.
\newline\ To establish the existence of metastable states,
Barlovic \textit{et al} started their simulations of the
VDR model from two different initial configurations, the megajam
and the homogeneous state. The megajam consists of one large
compact cluster of standing cars. In the homogeneous state, cars
are distributed periodically with equal constant gap between
successive cars (with one lager gap for incommensurate densities).
If the initial configuration is homogeneous, one obtains the upper
branch, for some interval of densities $[\rho_{1},\rho_{2}]$, in
which each car moves freely with maximal velocity (see figure 1).
This upper branch is metastable with an extremely long life-time.
If the initial configuration is megajam, one obtains the lower
branch which is phase separated. The phase separated state
consists of a large jam (jammed region) and a free-flow region
where each car moves freely. It is known that the lifetime of the
metastable states depends on the system length L. Yet, the
simulation results indicate that $\Delta \rho = \rho_{1} -
\rho_{2}$ decreases with larger system sizes and is expected to
vanish for $L\rightarrow\infty$, i.e. the jammed branch is stable
in that limit. Therefore the non-unique behaviour of the
fundamental diagram is only observable if finite system sizes are
considered (see \cite{chow} page 93).
\newline\ As vehicular traffic usually evolved in multi-lane roads, some
interesting question is not yet studied. Does multi-lane version
of the VDR model always exhibits metastable states?
\newline\ The NaSch model with VDR rule is a one-dimensional
probabilistic CA which consists of $N$ cars moving on a
one-dimensional lattice of $L$ cells with periodic boundary
conditions (the number of vehicles is conserved). Each cell is
either empty, or occupied by just one vehicle with velocity
$v=1,2,...,v_{\max }$. We denote by $x_{k}$ and $v_{k}$ the
position and the velocity of the {\it kth} car at time $t$
respectively. The number of empty cells in front of the {\it kth}
car is denoted by $d_{k} =x_{k+1} -x_{k} -1$ and called hereafter
as the gap. Space and time are discrete. At each discrete
time-step $t\rightarrow t+1$ the system update is performed in
parallel for all cars according to the following four subrules :
\newline\ \textbf{$R_{1}$}: VDR, $p(v_{k})=p_{0}$ for $v_{k}=0$
and $p(v_{k})=p$ for $v_{k}>0$.
\newline\ \textbf{$R_{2}$}: Acceleration,
$v_{k}\leftarrow \min \left( v_{k}+1,v_{\max }\right)$ .
\newline\ \textbf{$R_{3}$}: Slowing down, $v_{k}\leftarrow \min \left(
v_{k},d_{k}\right)$.
\newline\ \textbf{$R_{4}$}: Randomization, $
v_{k}\leftarrow \max \left( v_{k}-1,0\right) $ with probability
$p(v_{k})$.
\newline\ \textbf{$R_{5}$}: Motion, the car is moved
forward according to its new velocity, $x_{k}\leftarrow
x_{k}+v_{k}$.
\newline\ In two-lane traffic models, lane changing of vehicles
are performed according to some additional rules [5-9]. In this
paper, we shall adopt the symmetric exchange rules which are
defined by the following criteria \cite{Chow2}:
\begin{enumerate}
  \item $min(v_{k}+1,v_{max})>d_{k}$
    \item $d_{k,other}>d_{k}$ and $d_{k,back}> l_{back}$
   \item $p_{ch}>rand()$
\end{enumerate}
Here $d_{k,other}$ (resp. $d_{k,back}$) denotes the gap on the
target lane in front of (resp. behind) the car that wants to
change lanes. Two different formulas are assigned to the parameter
$l_{back}$. For aggressive drivers, i.e., vehicles which do not
look back before changing lanes, we choose $l_{back}=0$. For
careful drivers, i.e., vehicles which respect the safety
criterion, we choose $l_{back}=v^{b}_{o}+1$, where $v^{b}_{o}$ is
the velocity of the following car in the target lane. Finally,
$p_{ch}$ is the lane-changing probability and $rand()$ stands for
a random number between 0 and 1. Hereafter, we shall denote by
VDRM$_{N_{a}}$ the two-lane VDR traffic models where $N_{a}$
represents the number of aggressive drivers present in the
circuit.
\newline\ The update in the two-lane model is divided into two
sub-steps: in one sub-step, cars may change lanes in parallel
following the above lane changing rules and in the other sub-step
each car may move effectively by the forward movement rules as in
the single-lane traffic.
\newline\ We performed computer simulations of the two-lane model
with the following parameters, ($p_{0}=0.01$, $p=0.7$ and
$v_{max}=5$). The size of the lattice is given by $L=1000$.
Starting from an initial configuration (homogeneous or megajam)
the system evolved in time steps with respect to the above
dynamical rules. For each simulation run, we discarded some number
($t_{dis}$) of time steps and we performed averages of the flow
over $t_{av}=50 000$ time steps. The duration of each run is
"$t_{dis}+t_{av}$". The procedure is then repeated for a number
$100$ of different realizations of the homogeneous (or megajam)
initial configurations. The average over all the different
realizations gives a mean value of the flow.
\newline\ Figure 1 illustrated the variation of the flow $J$, in
two-lane VDR traffic models, as a function of the density of cars
and for different values of the discarded time $t_{dis}$. We
noticed that the flow in both lanes are equal since symmetric lane
changing are considered. First, we shall consider the case where
only one aggressive driver is present in the circuit. So, if the
homogeneous initial state is used, a higher branch of the flow is
observed for some interval of densities $[\rho_{1},\rho_{2}]$
whenever $t_{dis}$ is small enough. When increasing enough
$t_{dis}$, the high branch interval diminished and disappeared
completely at certain limit of $t_{dis}$. Notice that this
phenomena occurred also for the NS model with very small
randomization $p$. In contrast, the upper branch in the
fundamental diagram of the VDRM$_{0}$ does not change when one
increases the time $t_{dis}$ (Fig. 1). This shows clearly that the
hysteresis exist in the fundamental diagram of the VDRM$_{0}$
model but not in the one of the VDRM$_{1}$.
\newline\ To clarify more the above results, we shall consider the
time evolution of the flow for some fixed density $\rho=0.12$
($\rho_{1}<\rho<\rho_{2}$) and for the homogeneous and megajam
initial states (Fig. 2). It is shown that in contrast to the
VDRM$_{0}$, where the homogeneous state is metastable with an
extremely long life-time, this state does not exist in VDRM$_{1}$.
Yet, in this later, the flow corresponding to the homogeneous
initial configuration decreases with time until reaching the value
corresponding to the megajam initial configuration. The breakdown
of the homogeneous structure in the two lanes is due to the
occurrence of stopped cars provoked by the abrupt lane changing of
the aggressive driver. Figure 3 shows the evolution of the density
of stopped cars in the lanes when starting from the initial
homogeneous state. In VDRM$_{0}$, no stopped cars exist in the
circuit because all cars respect the safety criteria of lane
changing. However, in VDRM$_{1}$, the density increases with time
until it reaches a stationary value. Stopped cars act as
perturbations for the free flow region and as such trigger the
breakdown of the high flow states.
\newline\ In figure 4, we show the cluster size distribution in the steady
state of VDRM$_{1}$ for different lane-changing probability
$p_{ch}$. The cluster means here a string of successive stopped
cars in a single lane of the two-lane model, i.e. we are
considering only compact jams in a single lane. As the symmetric
lane-changing rules are considered here, the cluster sizes
distribution in the two lanes must be equal. We observe from
figure 4 the bimodal nature of the cluster size distribution as
$p_{ch}\simeq1$. Large clusters appear in the lanes but there are
by far many more small-sized clusters than large ones.
Furthermore, with decreasing $p_{ch}$, the probability of small
clusters increases while that of large clusters diminishes. If
$p_{ch}=0$, which corresponds to the single lane VDR model, almost
all cars are congested in one large cluster with the exception of
a few isolated cars. This is the well known phase separated state.
\newline\ In this section, we shall investigate the relaxation
dynamics of VDRM$_{N_{a}}$ for different values of $N_{a}$ and
$p_{ch}$, when starting from the homogeneous initial condition.
This is done by plotting the time evolutions of the flow and
computing their relaxation times. Hence, the greater is the number
of aggressive drivers, the faster is the system relaxation (Fig.
5). When decreasing the probability of lane changing $p_{ch}$, one
sees that the equilibration is delayed. Indeed, in this case, the
abrupt lane changes of aggressive cars become less frequent and
the number of stopped cars becomes small (Fig. 6).
\newline\ To study numerically the relaxation time corresponding
 to an observable $A$ we shall use the nonlinear relaxation function \cite{Bind}:
\begin{equation}
\phi (t)=[A(t)-A(\infty )]/[A(0)-A(\infty )] \
\end{equation}
The corresponding nonlinear relaxation time
\begin{equation}
\tau =\int_0^\infty \phi (t)dt.  \
\end{equation}
The condition that the system is well equilibrated is
\begin{equation}
t_{M_0}\gg \tau   \
\end{equation}
where $M_0$ is the number of Monte Carlo steps that have to be
excluded in the averaging of the observable $A$. In figure 7, we
plotted the variation of the relaxation time $\tau$ of the
observable $J$ near the limit $p_{ch}\rightarrow 0$. As a result,
the relaxation time is found to diverge as $p_{ch}\rightarrow 0$.
Moreover, we see that $\tau$ follows a power law behavior of the
form,
\begin{equation}\
    \tau\propto p_{ch}^{-\beta}
\end{equation}
Except for some minor fluctuations, the dynamic exponent $\beta$
remains unchanged when varying the number of aggressive drivers
present in the circuit. For example, $\beta\approx
0,9798\pm0,0290$ for $N_{a}=1$ and $\beta\approx 0,9801\pm0,0315$
for $N_{a}=3$. Assuming that the parameter $p_{ch}$ is rate of
transition for the dynamics of the model, and as it was
demonstrated in Ref. \cite{mou2}, the exponent $\beta$ is expected
to be theoretically equal to one.
\newline\
In summary, we have shown that the presence of aggressive drivers
in the circuit breakdowns the state of high traffic flow. In
theses systems neither phase separation nor metastability can
occur and a new stationary state takes place. Indeed, the abrupt
lane changing of aggressive drivers force the succeeding cars on
the destination lane to decelerate enough; leading therefore to
the occurrence of stopped cars and then the formation of jams.
\newline\ In the NS model, the cluster sizes distribution decreases exponentially
while in VDRM$_{0}$, it should depend on the initial state. Yet,
for some density $\rho$ in the hysteresis region
($\rho_{1}<\rho<\rho_{2}$), no clusters appear in the metastable
homogeneous state. However, a big cluster persists in the phase
separated state. Nevertheless, in VDRM$_{N_{a}}$ where
$N_{a}\neq0$, the stationary state is composed by small and big
clusters. The distribution of these clusters should depend on the
lane-changing probability $p_{ch}$. As $p_{ch}$ decreases the
relaxation time $\tau$ of the system increases and diverges at the
limit $p_{ch}\rightarrow 0$. The relaxation behaviour follows a
power law behavior of the form, $\tau\propto p_{ch}^{-\beta},
\beta=1$).
\newpage\

\newpage\ \textbf{Figures captions}
\begin{quote}
\textbf{Figure 1}. Illustration of the variation of the flow $J$
in the two-lane VDR traffic models as a function of the density of
cars and for different values of the discarded time $t_{dis}$
($p_{ch}=0.10$).
\newline\ \textbf{Figure 2}. Time evolutions of the flow for homogeneous and megajam
initial states ($\rho=0.12$ and $p_{ch}=0.10$).
\newline\ \textbf{Figure 3}. Time evolution of the density of
stopped cars when starting from the initial homogeneous state
($\rho=0.12$, $N_{a}=1$ and $p_{ch}=0.10$).
\newline\ \textbf{Figure 4}. The cluster size
distribution in the steady state for different lane-changing
probability $p_{ch}$ ($\rho=0.12$ and $N_{a}=1$).
\newline\ \textbf{Figure 5}. Time evolutions of
the flow and the density of stopped cars, when starting from
homogeneous initial configuration, for several values of $N_{a}$
($\rho=0.12$ and $p_{ch}=0.10$).
\newline\ \textbf{Figure 6}. Time evolutions of
the flow and the density of stopped cars, when starting from
homogeneous initial configuration, for several values of $p_{ch}$
($\rho=0.12$ and $N_{a}=1$).
\newline\ \textbf{Figure 7}. Variations of the relaxation time $\tau$
 near the limit $p_{ch}\rightarrow 0$ ($\rho=0.12$).
\end{quote}
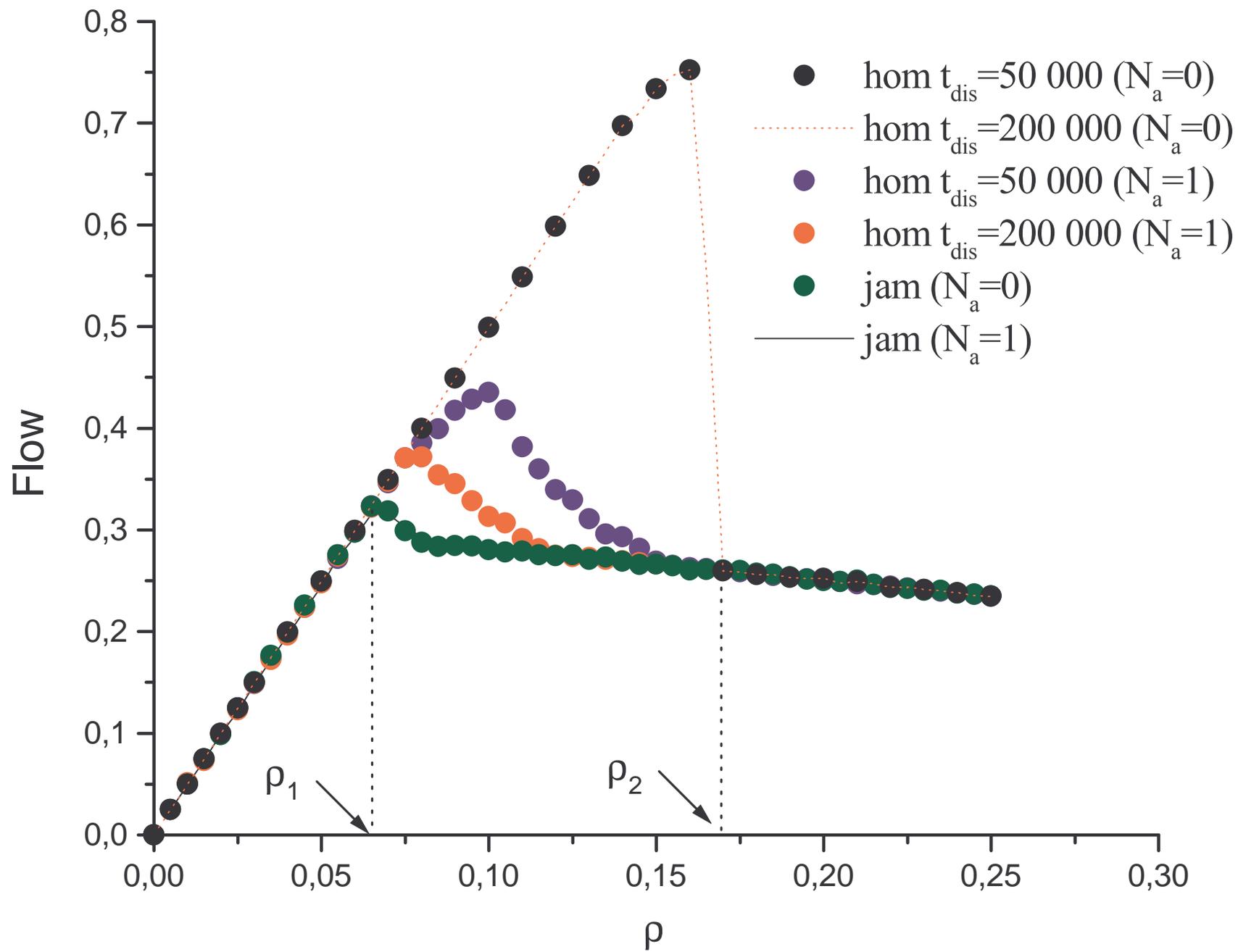
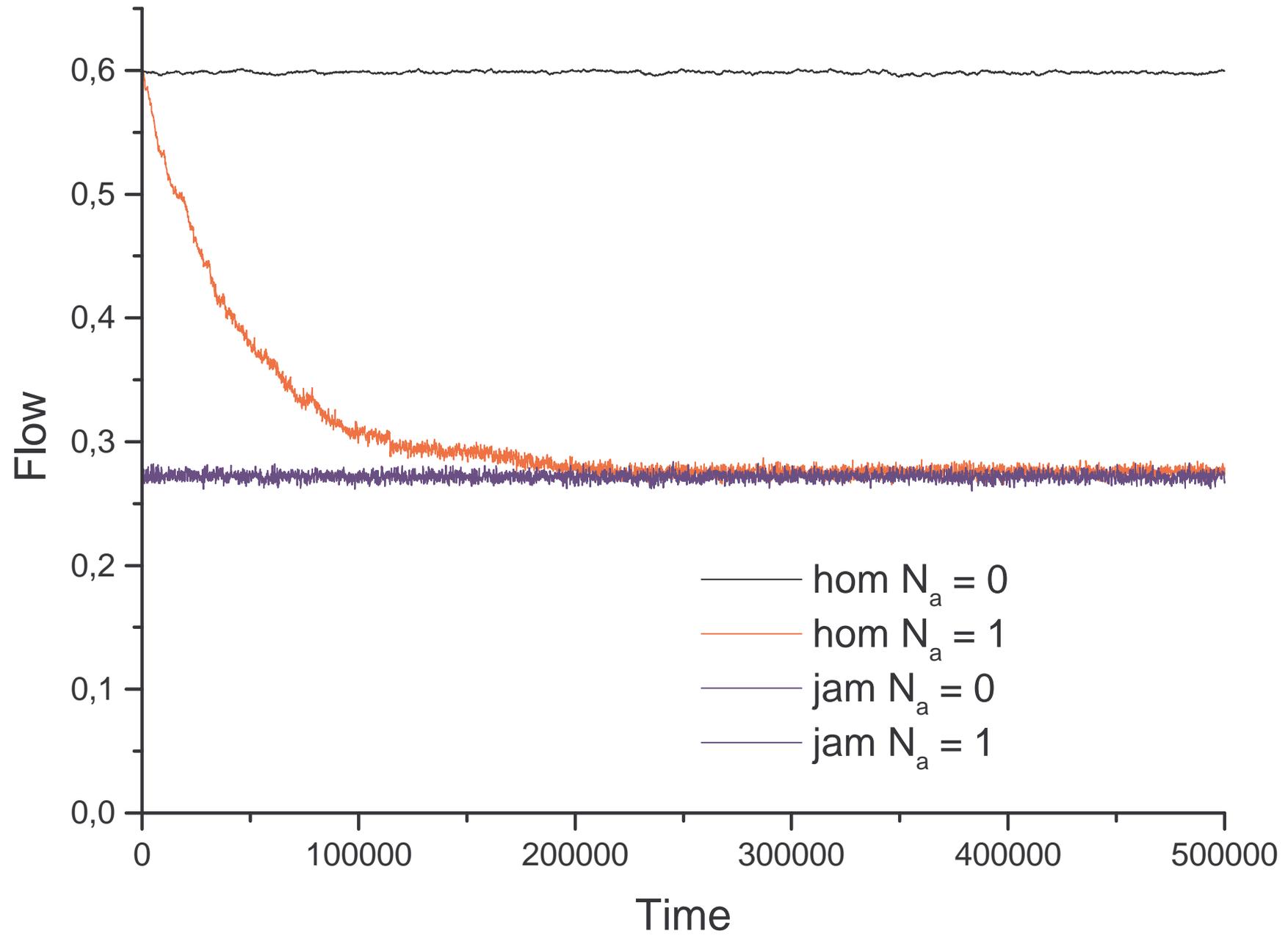
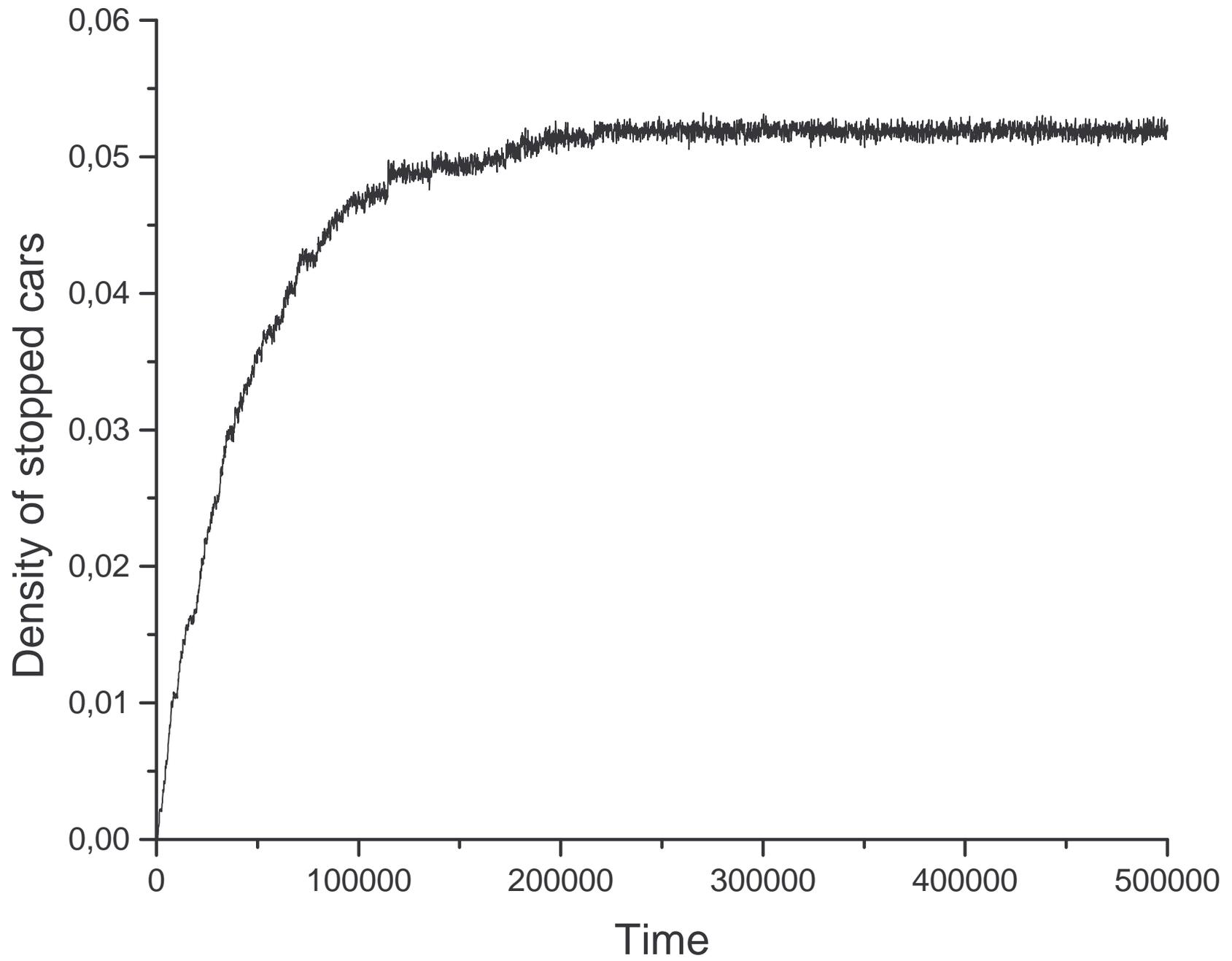
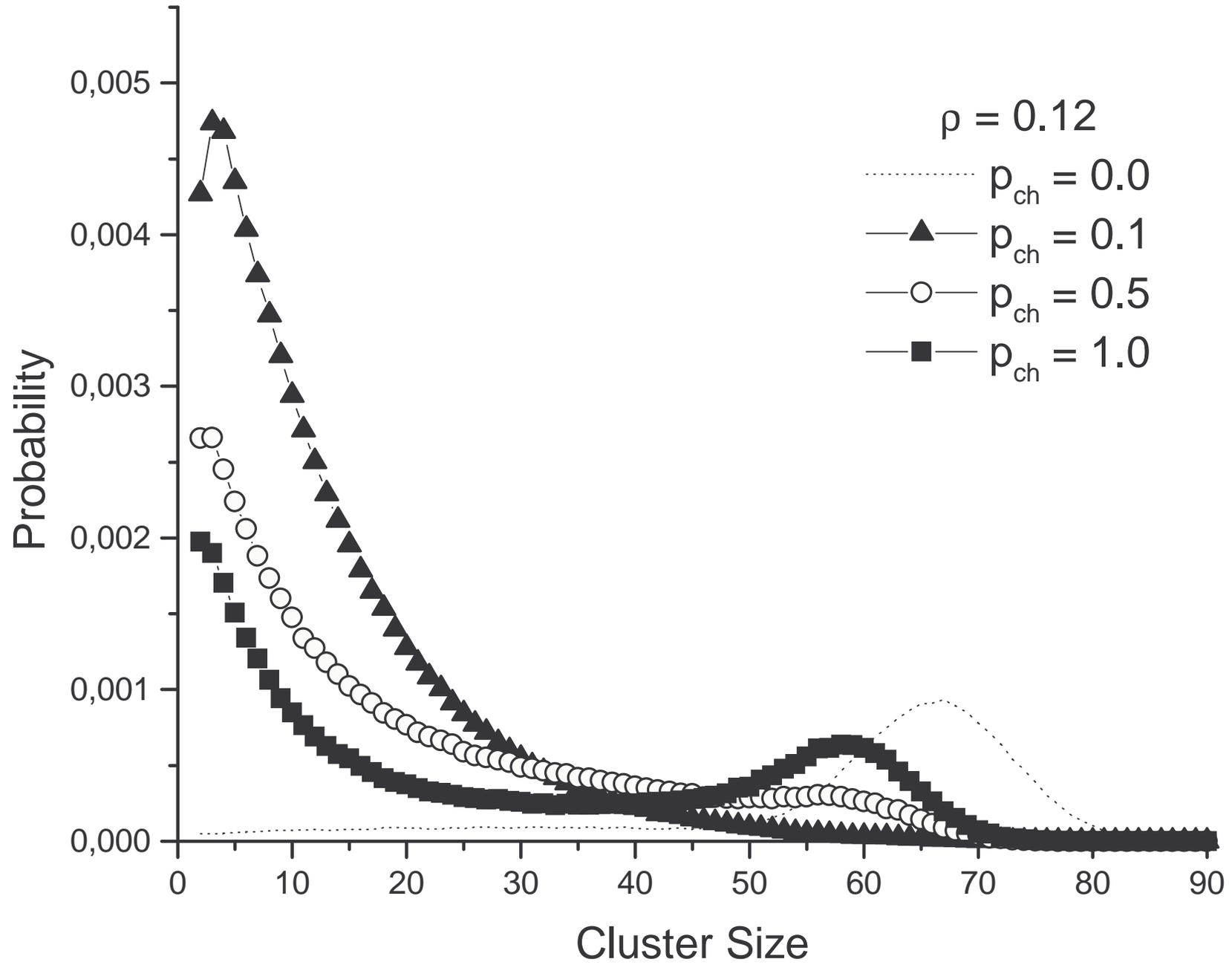
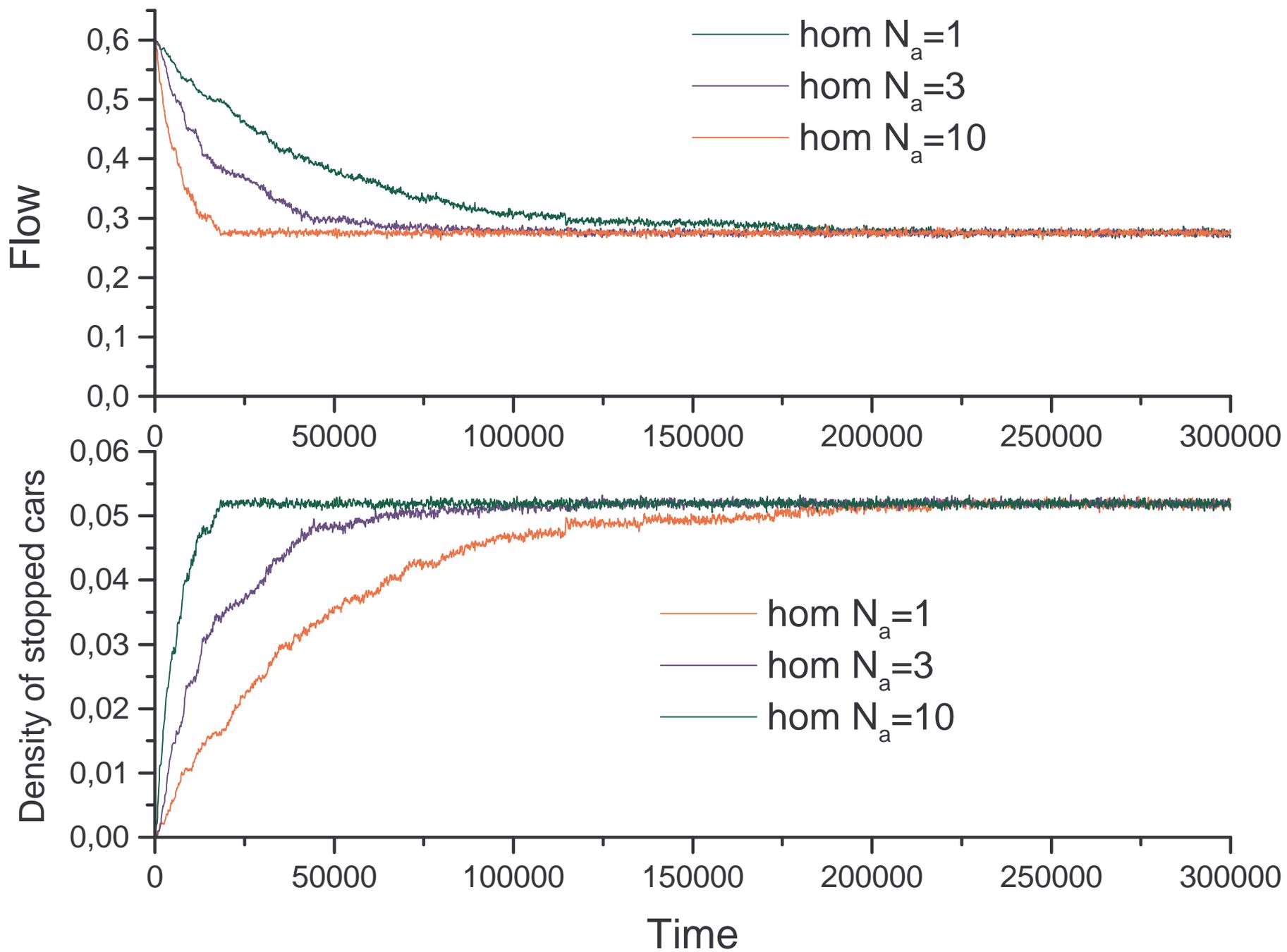
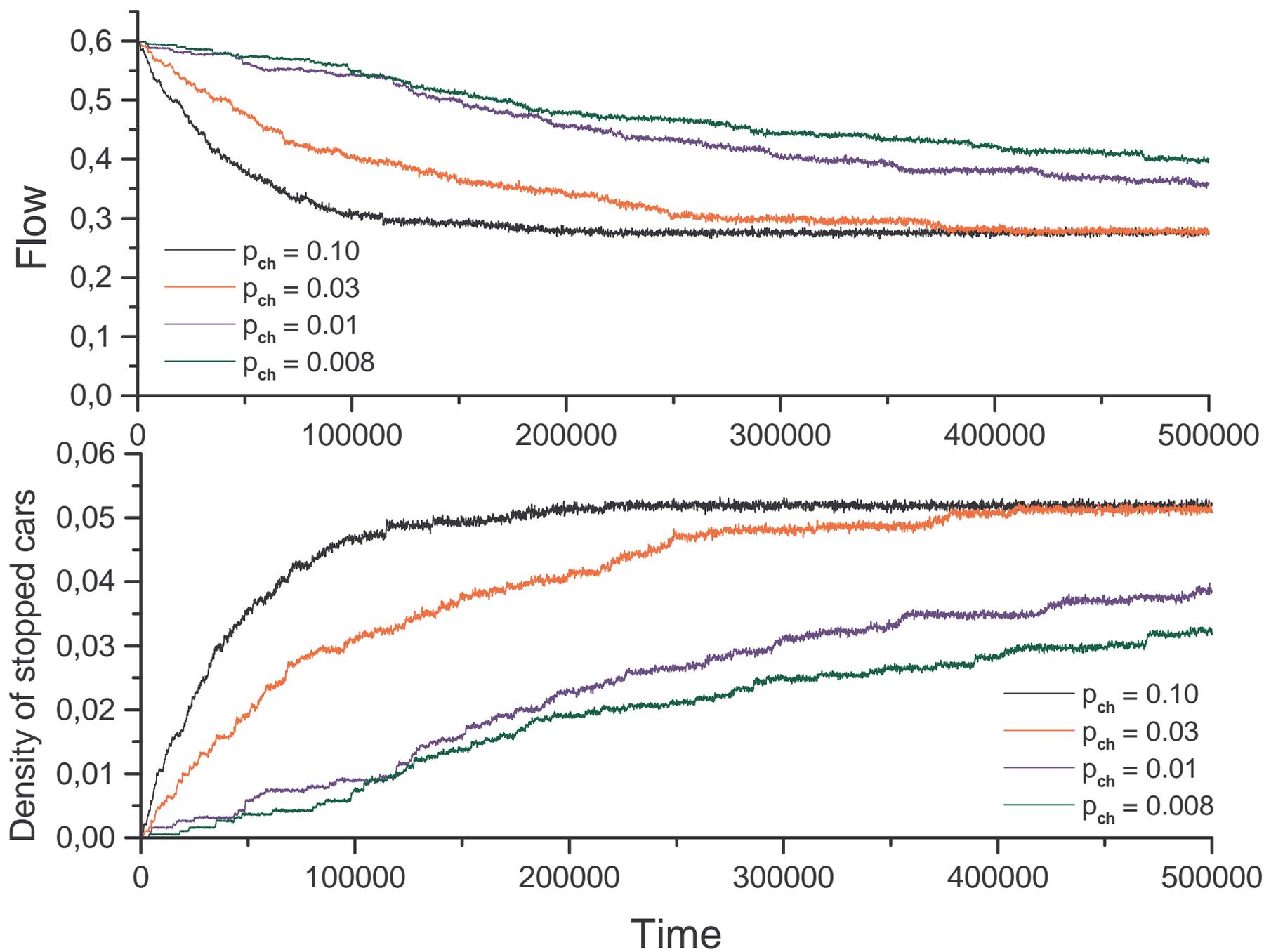
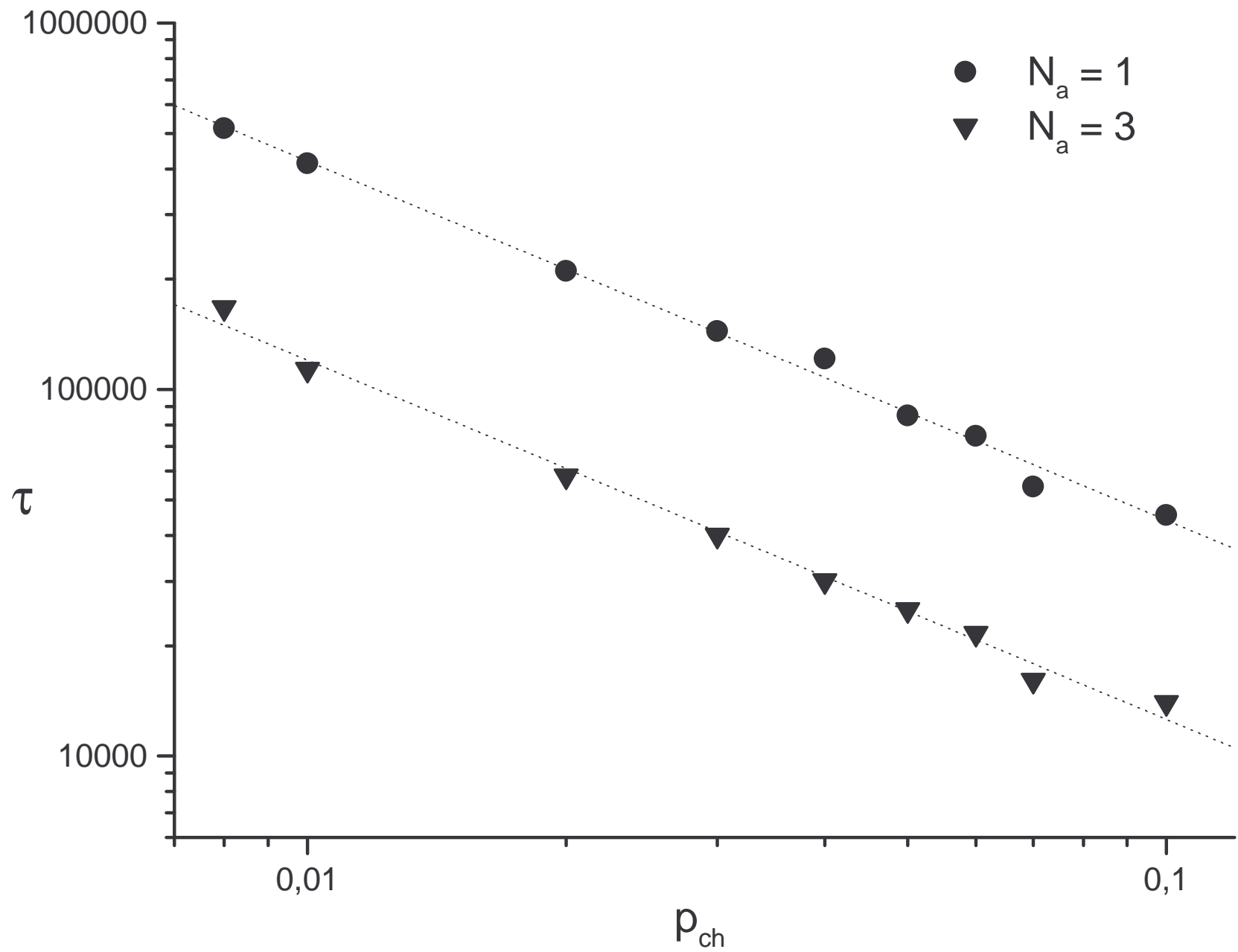
\end{document}